\documentclass[original]{adada}

\usepackage{graphicx}
\usepackage[T1]{fontenc}
\usepackage{amsmath,amssymb}
\usepackage{booktabs}
\usepackage{caption}
\usepackage{float}
\usepackage{url}
\usepackage{stfloats}
\usepackage{cuted}
\usepackage{tikz}
\usepackage{pgfplots}
\pgfplotsset{compat=1.18}
\usetikzlibrary{arrows.meta,positioning,fit,calc,backgrounds}
\DeclareCaptionLabelSeparator{adadaspace}{\space}
\makeatletter
\let\oldthebibliography\thebibliography
\renewcommand{\thebibliography}[1]{\oldthebibliography{#1}\setlength{\itemsep}{\baselineskip}\setlength{\parskip}{0pt}}
\makeatother

\title{Beyond Generation: An Empirical Study on Redefining the Act of Drawing Through an 85\% Time Reduction in Picture-Book Production}
\author{Cosei Kawa}{Faculty of Design, Shizuoka University of Art and Culture, Shizuoka, Japan}{k-cosei@suac.ac.jp}

\begin{document}

\begin{abstract}
Conventional picture-book production imposes substantial physical and temporal demands on creators, often constraining opportunities for high-level artistic exploration. While generative AI can drastically accelerate image generation, concerns remain regarding style homogenization and the erosion of authorial agency in professional practice. This study presents an empirical evaluation of an AI-collaborative workflow through the full production of one professional 15-illustration picture-book title, and compares the process with a conventional hand-drawn pipeline by the same creator. Quantitatively, the proposed workflow reduces total production time by 85.2\% (from 2,162.8 to 320.4 hours), with the largest substitution observed in early drafting stages. Qualitatively, however, the core contribution is the strategic reallocation of labor: time saved in mechanical rendering is reinvested into high-level \textit{Judgment} (aesthetic selection, narrative direction, and cross-scene consistency decisions) and \textit{Completion} (embodied manual retouching and integrative refinement). Notably, 235 hours were devoted to Completion, indicating that publication-quality outcomes still depend on sustained human synthesis to resolve emotional nuance, stylistic coherence, and narrative continuity beyond current automated generation. We therefore frame human-AI collaboration not as replacement, but as a human-centered restructuring of creative work in which AI increases exploratory throughput while human authorship remains decisive in final expressive value. This evidence supports a practical model for balancing measurable efficiency with domain-specific artistic responsibility in professional storytelling production.
\end{abstract}


\maketitle

\begin{strip}
  \centering
  \captionsetup{width=0.96\textwidth}
  \includegraphics[width=0.98\textwidth]{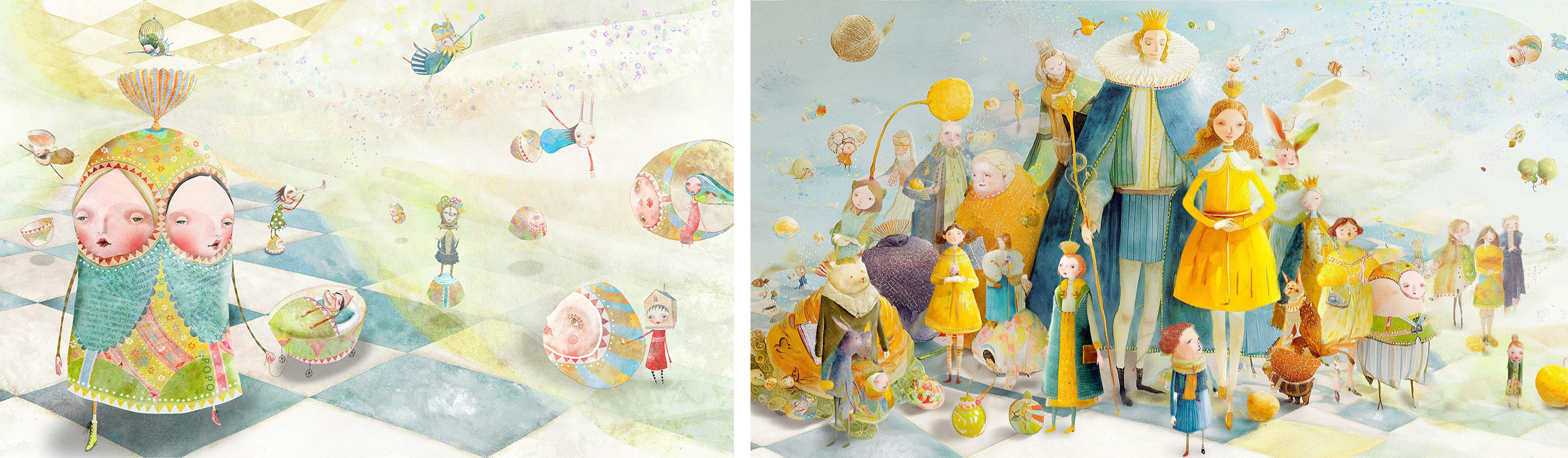}
  \captionof{figure}{Comparison of conventional and AI-collaborative picture-book illustration workflows.
  (a) \textit{Carpe Diem}, produced with a conventional hand-drawn pipeline (1,203.6 hours for illustration).
  (b) \textit{Golden Drops Opening the Sky}, produced with our AI-collaborative workflow (1.2 hours for initial AI output + 235.3 hours for human Completion).
  The proposed method inherits and further develops picture-book-specific stylistic signatures from the author’s prior works.
  More importantly, by reinvesting recovered time into aesthetic selection and scene direction, the workflow yields richer narrative cues in props and architecture, as well as a stronger sense of atmospheric presence.
  Reported times indicate total effort per one picture-book title (15 illustrations). All illustrations \copyright Cosei Kawa. All rights reserved.}
  \label{fig:teaser}
  \vspace{\baselineskip}
\end{strip}

\section{Introduction}
Generative Artificial Intelligence (AI) is rapidly reshaping visual design workflows \cite{epstein2023, oppenlaender2022}. Although current systems can produce high-fidelity images at high speed, HCI research continues to warn against style homogenization and reduced human agency \cite{shneiderman2022}. This tension is especially critical in picture-book production, where narrative continuity, stylistic coherence, and emotional resonance are central and cannot be guaranteed by autonomous generation alone.

Despite the growth of co-creative systems, empirical studies in real professional pipelines remain limited, particularly for post-generation, embodied refinement. To address this gap, we evaluate an AI-collaborative workflow in the production of a 15-illustration professional picture book (see Figure~\ref{fig:teaser}), asking how generative AI can amplify rather than replace authorial practice.

\begin{figure}[H]
    \centering
  \includegraphics[width=0.88\columnwidth]{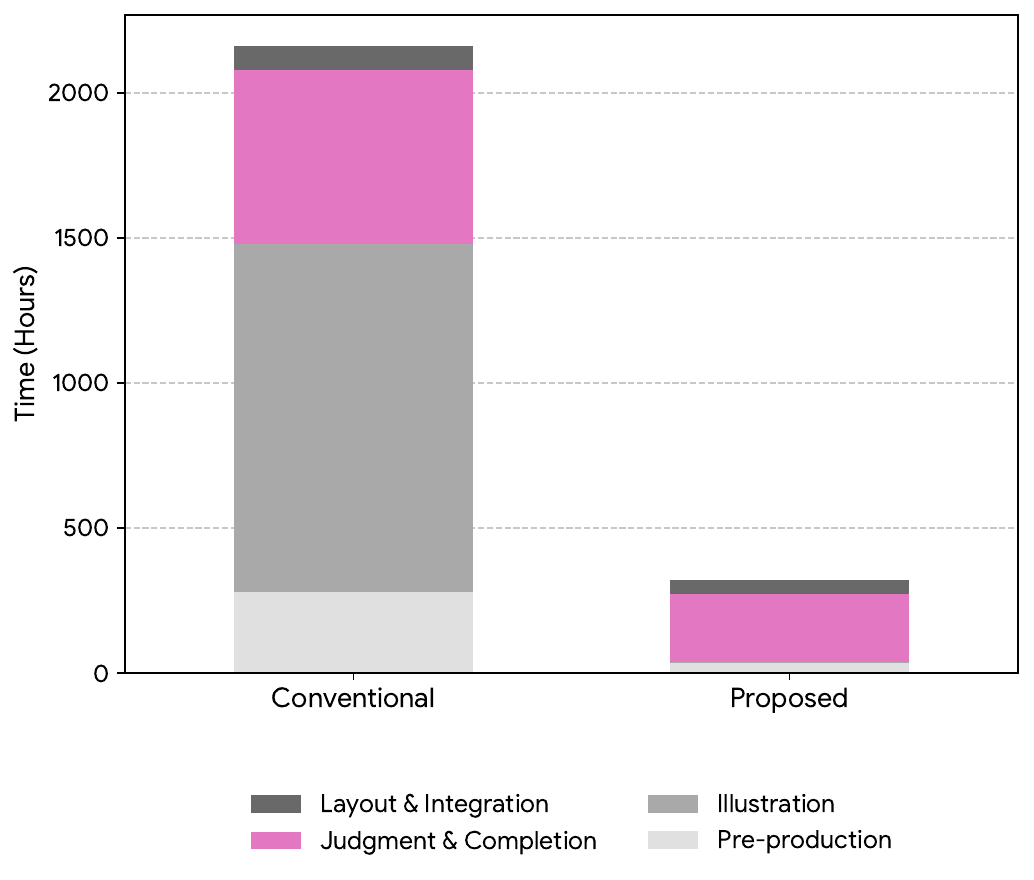}
  \caption{Comparison of production time breakdown between conventional and proposed workflows. While total time is reduced by 85\%, the proportion of time dedicated to 'Judgment \& Completion' significantly increases in the proposed method.}
  \label{fig:time_graph}
\end{figure}

The key benefits of the proposed method are summarized as follows:
\begin{itemize}
  \setlength{\itemsep}{0.5\baselineskip}
  \item \textbf{Large efficiency gain:} total production time is reduced by 85\%, from 2,162.8 hours to 320.4 hours.
  \item \textbf{Role redefinition of the creator:} effort shifts from mechanical line execution to high-level \textit{Judgment} and \textit{Completion}.
  \item \textbf{Quality-focused time reinvestment:} 235 hours are deliberately reinvested in manual Completion to improve emotional resonance and narrative coherence.
  \item \textbf{Human-centered collaboration:} AI functions as a high-throughput drafting instrument while human authorial agency remains central.
\end{itemize}
\paragraph{Supplemental Materials} Video demonstrations of the ``Prompt Recipe'' UI and the final animated picture-book, ``Golden Drops Opening the Sky,'' are available on YouTube: \url{https://youtu.be/i8eorQ8SGQs} (UI Workflow) and \url{https://youtu.be/MAoIRbUYLe8} (Final Artwork). The physical publication ``Carpe Diem'' is available at: \url{https://www.amazon.co.jp/dp/4600002245}.

\medskip
\section{Related Work}

\subsection{Human-AI Collaboration and Agency}
The integration of artificial intelligence into complex workflows has been fundamentally shaped by the concept of Human-Centered AI (HCAI) \cite{shneiderman2022}, which emphasizes reliable, safe, and trustworthy systems that amplify human performance. Foundational frameworks \cite{amershi2019} have proposed guidelines for effective human-AI interaction, highlighting the necessity of maintaining user control and transparency. However, as AI systems transition from analytical tools to generative partners, preserving human agency---particularly the intuitive and embodied decision-making processes inherent in artistic creation---remains a critical challenge.

\subsection{Generative AI in Creative Practices}
In the specific context of visual design and creative production, recent studies have explored the paradigm of co-creation between humans and generative AI \cite{liu2022, oppenlaender2022}. These investigations have identified profound opportunities for expanding design exploration, while simultaneously raising concerns regarding the homogenization of artistic style and the potential deskilling of human creators \cite{epstein2023}. Much of the existing HCI discourse focuses heavily on the "interaction phase" (e.g., prompt engineering, iterative generation) and how creators steer the AI's output.

\subsection{Embodied Practice and Picture-Book Affordances}
Beyond HCI, foundational design theories emphasize that creative professionals "think in action" through continuous physical dialogue with their materials \cite{schon1983, pallasmaa2009}. In the specific domain of picture-book production, visual elements are not merely decorative but carry profound narrative affordances \cite{nodelman1988, nikolajeva2001, shulevitz1997}. However, existing AI-collaborative systems rarely address how to support the intricate, embodied manual retouching required to satisfy these domain-specific narrative demands \cite{mccullough1998}.

\subsection{Identifying the Gap: The Value of "Completion"}
Despite the rich body of literature on co-creative systems, empirical evaluations focusing on the post-generation phase---specifically, the manual intervention required to elevate an AI-generated draft to a professional, publication-ready artwork---are notably sparse. Existing frameworks often overlook the labor-intensive but artistically crucial process of what we term "Completion." This study addresses this gap by presenting a quantitative and qualitative analysis of an actual picture-book production pipeline. By demonstrating an 85\% reduction in overall production time, we re-center the academic conversation on how the time reclaimed from initial illustration can be strategically reinvested into human aesthetic Judgment and physical retouching, thereby preserving the author's unique stylistic signature.

\medskip
\section{Methodology}
\subsection{Reproducibility Setup}
To support reproducibility, we specify the operational stack used in this case study. Image generation was conducted with Midjourney v6 and parameterized style references; textual and metadata assistance was handled by LLM services including Gemini for automated tag extraction. The custom \textit{Prompt Recipe} environment was implemented in Python and Streamlit, with asynchronous dispatch utilities and metadata-writing scripts integrated into one production interface.

The production corpus consisted of approximately 200 prior artworks created by the same author, which were used as controlled style-reference assets rather than as a public benchmark dataset. All timing analyses reported in this paper were measured at the level of one complete picture-book title (15 illustrations), and development time for the software itself was excluded from production accounting. This separation between tool-construction effort and art-production effort is intended to make replication of the evaluation protocol transparent.

The reported process times in this validation (conventional workflow: 2,162.8 hours; proposed AI-collaborative workflow: 320.4 hours) are not rough estimates but aggregated \textit{Pure Working Hours} derived from the author's detailed production notes and task logs. Specifically, we extracted only intervals of direct engagement with physical and cognitive production tasks, excluding breaks and non-task contemplation. As the log data indicate, the most pronounced redistribution of time appears in the dramatic compression of the Illustration (main rendering) stage (1,203.6 hours to 1.2 hours) and the concentration of effort in the Completion (manual retouching and correction) stage (235.3 hours).

\subsection{Custom System Architecture for Creative Collaboration}
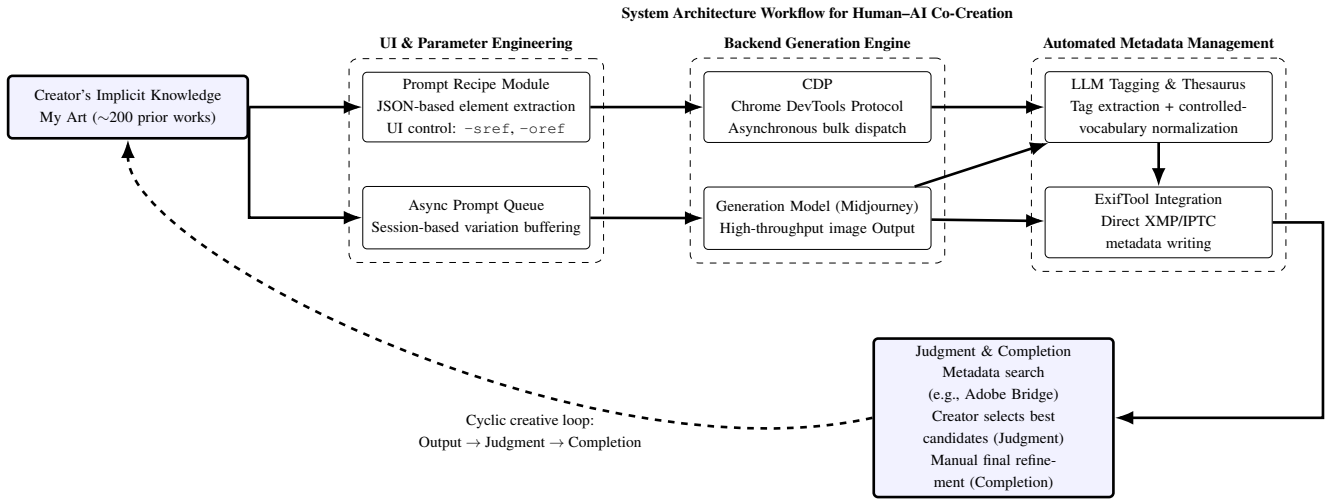
\begin{figure*}[!t]
\centering
\resizebox{1\textwidth}{!}{%
\begin{tikzpicture}[
    font=\footnotesize,
    >=Latex,
    node distance=9mm and 10mm,
    box/.style={draw, rounded corners=2pt, align=center, minimum height=10mm, inner sep=3pt},
    major/.style={box, very thick, fill=blue!5, text width=36mm},
    module/.style={box, fill=white, text width=34mm},
    data/.style={box, fill=green!6},
    flow/.style={->, very thick},
    feedback/.style={->, very thick, dashed},
    group/.style={draw, dashed, rounded corners=4pt, inner sep=6pt}
]

\node[major] (implicit) {Creator's Implicit Knowledge\\\footnotesize My Art (\(\sim\)200 prior works)};
\node[module, right=18mm of implicit] (prompt) {Prompt Recipe Module\\\footnotesize JSON-based element extraction\\\footnotesize UI control: \texttt{--sref}, \texttt{--oref}};
\node[module, below=7mm of prompt] (queue) {Async Prompt Queue\\\footnotesize Session-based variation buffering};

\node[module, right=18mm of prompt] (cdp) {CDP\\\footnotesize Chrome DevTools Protocol\\\footnotesize Asynchronous bulk dispatch};
\node[module, below=7mm of cdp] (gen) {Generation Model (Midjourney)\\\footnotesize High-throughput image Output};

\node[module, right=18mm of cdp] (tagging) {LLM Tagging \& Thesaurus\\\footnotesize Tag extraction + controlled-vocabulary normalization};
\node[module, below=7mm of tagging] (exif) {ExifTool Integration\\\footnotesize Direct XMP/IPTC metadata writing};

\node[major, below=14mm of gen, xshift=28mm] (jc) {Judgment \& Completion\\\footnotesize Metadata search (e.g., Adobe Bridge)\\\footnotesize Creator selects best candidates (Judgment)\\\footnotesize Manual final refinement (Completion)};

\draw[flow] (implicit) -- (prompt);
\draw[flow] (implicit.east) |- (queue.west);

\draw[flow] (prompt) -- (cdp);
\draw[flow] (queue) -- (gen);

\draw[flow] (cdp) -- (tagging);
\draw[flow] (gen) -- (exif);
\draw[flow] (gen) -- (tagging);
\draw[flow] (tagging) -- (exif);

\draw[flow] (exif.east) -- ++(8mm,0) |- (jc.east);

\draw[feedback] (jc.west) .. controls +(-45mm,-10mm) and +(0mm,-18mm) .. (implicit.south)
    node[pos=0.30, below, align=center, xshift=-9mm] {\footnotesize Cyclic creative loop:\\\footnotesize Output \(\rightarrow\) Judgment \(\rightarrow\) Completion};

\begin{scope}[on background layer]
    \node[group, fit=(prompt)(queue),
        label={[align=center]north:\textbf{UI \& Parameter Engineering}}] {};
    \node[group, fit=(cdp)(gen),
        label={[align=center]north:\textbf{Backend Generation Engine}}] {};
    \node[group, fit=(tagging)(exif),
        label={[align=center]north:\textbf{Automated Metadata Management}}] {};
\end{scope}

\node[align=center, font=\bfseries\small, above=7mm of $(prompt.north)!0.5!(tagging.north)$] (title)
{System Architecture Workflow for Human--AI Co-Creation};

\end{tikzpicture}
}
\caption{
System architecture workflow of our human--AI collaborative picture-book pipeline.
The architecture operationalizes role separation: AI handles rapid provisional \textit{Output},
while the creator reinvests recovered time in \textit{Judgment} (aesthetic selection) and
\textit{Completion} (manual final integration). Dashed module boundaries indicate technical layers
that guarantee this division of labor across UI-level parameter engineering, backend asynchronous
generation, and automated metadata management.
}
\label{fig:system_arch}
\end{figure*}

Our framework integrates LLMs and image-generation models (e.g., Midjourney), grounded in latent diffusion-based generation paradigms \cite{rombach2022}, with an in-house platform (Python + Streamlit) designed to preserve creator agency.
To address this, our proposed workflow adopts a co-creative paradigm \cite{amershi2019, shneiderman2022}. Rather than treating the AI as an autonomous creator, we utilize it as a high-throughput drafting tool, strictly adhering to principles that emphasize user control \cite{amershi2019}.
The end-to-end architecture is summarized (see Figure~\ref{fig:system_arch}), and the corresponding production interface is shown (see Figure~\ref{fig:ui_screenshot}).


The concrete workflow is organized into three modules:

\textbf{(1) Prompt Recipe Module:} analyzes the author’s prior works and parameterizes thematic, compositional, and chromatic elements; exposes style-reference controls (e.g., \texttt{--sref}, \texttt{--oref}) in a UI so tacit style knowledge becomes controllable and reusable.

\medskip

\textbf{(2) Async Prompt Queue + Backend Dispatch:} buffers large sets of parameter-perturbed prompts and dispatches them asynchronously (CDP-based automation), removing synchronous waiting and increasing exploratory throughput.

\medskip

\textbf{(3) Automated Metadata Management:} applies LLM-based tagging and thesaurus normalization, then writes metadata directly into files (XMP/IPTC via ExifTool) for scalable retrieval and curation.

\vspace{\baselineskip}
\begin{center}
  \centering
  \includegraphics[width=\columnwidth]{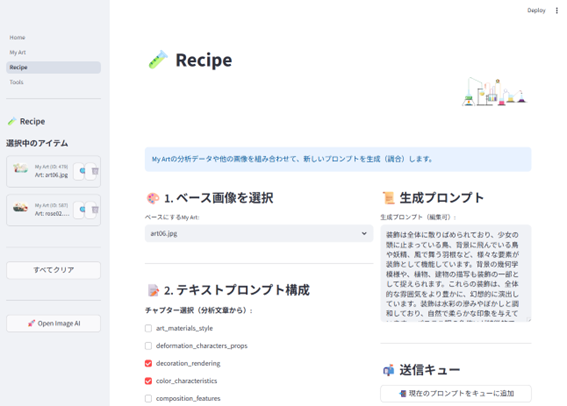}
  \captionof{figure}{Graphical User Interface of the 'Prompt Recipe' system. Note: The UI is in Japanese as it was natively developed for the author's production environment. (A: Style Slider, B: Prompt Queue for iterative generation based on the author's legacy assets.)}
  \label{fig:ui_screenshot}
\end{center}

\medskip
\begin{figure*}[!t]
  \centering
  \includegraphics[width=\textwidth]{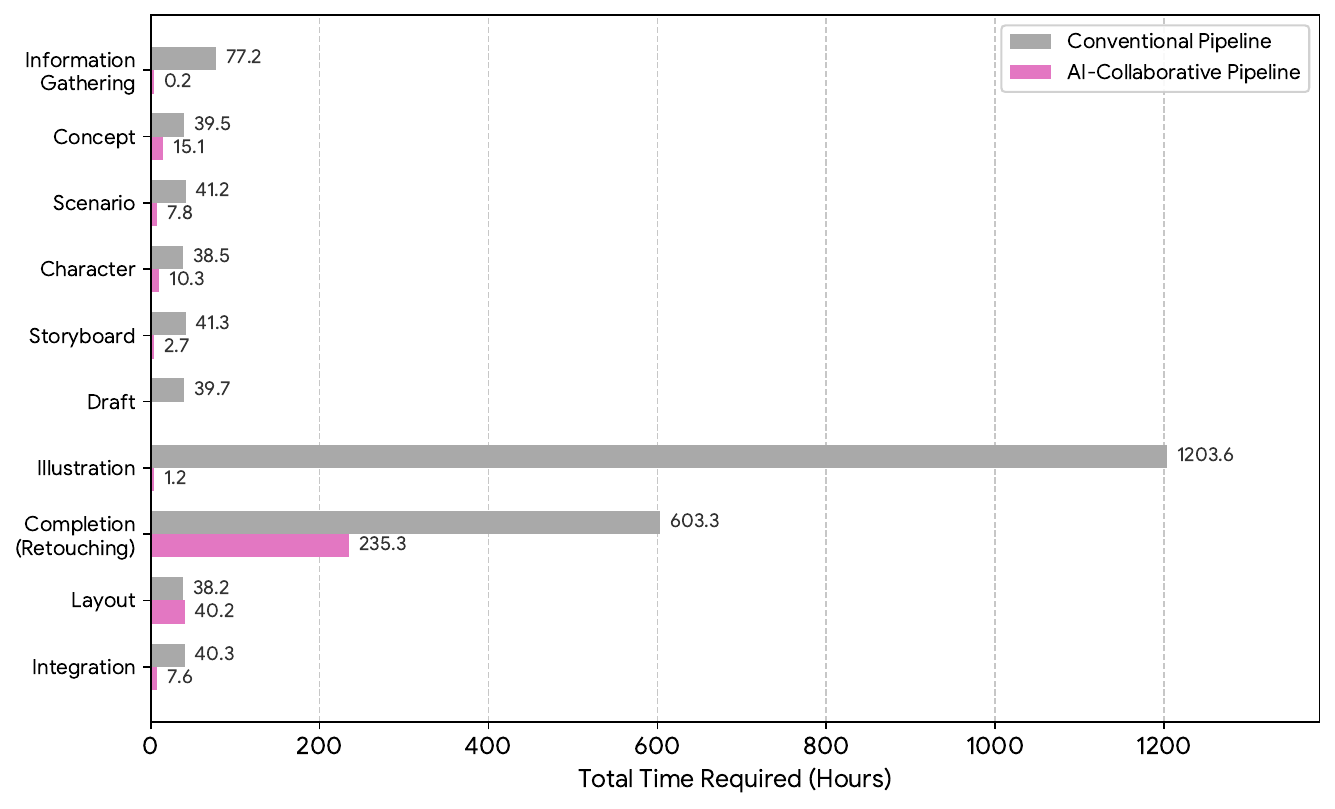}
  \caption{Total production time comparison per picture book (15 illustrations). The conventional hand-drawn pipeline is compared against the actual measured time of the AI-collaborative pipeline.}
  \label{fig:before_actual}
\end{figure*}

\subsection{Technical Implementation and Pipeline Automation}
From an HCI perspective, the central design goal was to externalize the creator's tacit style knowledge into a manipulable interface while preserving authorial control. To do so, we operationalized approximately 200 prior author-created works as style anchors in the \textit{Prompt Recipe} system. Rather than relying on free-text prompting alone, thematic and chromatic controls were exposed as explicit UI selections (e.g., checkbox-based recipe composition), and style-reference parameters such as \texttt{--sref} and \texttt{--oref} were numerically managed within the interface. This design transformed implicit craft memory into repeatable interaction primitives that could be reused across scenes.

The second implementation focus was Judgment support through asynchronous orchestration. During exploration, prompt variants with systematically perturbed controls (including \texttt{--chaos} and stylization values) were accumulated into a queue and dispatched in background batches. This removed synchronous waiting from the artist's interaction loop and reduced context switching between creative decision-making and machine operation. As a result, the creator could allocate attention to comparative evaluation across a large candidate set, which is the high-value Judgment process emphasized in co-creative HCI frameworks \cite{amershi2019, shneiderman2022}.

The third focus was the elimination of repetitive administrative noise. To reduce file-management burden, generated images were automatically processed by Gemini-based tag extraction, followed by thesaurus-based normalization to control lexical variation (e.g., synonym and notation inconsistencies). The normalized descriptors were then written directly into image assets as XMP/IPTC metadata via ExifTool, enabling robust retrieval and downstream curation without manual renaming or ad hoc folder maintenance.

Collectively, these automation layers did not aim to automate artistic intent itself; they reallocated interaction cost away from operational friction and toward aesthetic synthesis. In practical terms, the system shifted the creator's role from repeatedly issuing low-level commands to supervising style continuity, narrative fit, and emotional tone at a portfolio scale. This supports a human-centered interpretation of pipeline automation: the machine accelerates throughput, while the human retains final authority over expressive value.

\subsection{Iterative Prompting for Stylistic Coherence}
To generate the initial draft illustrations (the "Provisional Output"), we developed a highly constrained prompt engineering strategy designed to preserve the author's signature aesthetic. Rather than relying on open-ended text-to-image generation, which often results in stylistic volatility and unpredictable outputs \cite{liu2022, oppenlaender2022}, our approach utilized a combination of structured textual prompts and image-to-image conditioning. This aligns with recent findings that prompt engineering alone is insufficient for professional artistic control \cite{epstein2023}. We systematically optimized prompts to control character consistency, watercolor-like textural softness, and foundational compositional layouts. This structured generation phase minimized unpredictable visual hallucinations, ensuring that the raw AI outputs served as viable structural scaffolds rather than finished artworks, thereby streamlining the subsequent Judgment and Completion phases.

\subsection{Style Adaptation and Hidden Time}
To ensure stylistic consistency, the generation workflow is informed by approximately 200 prior artworks by the author.
Accordingly, the measured reduction in current production time should not be interpreted as a purely technical gain detached from prior artistic investment.
The method depends on accumulated tacit knowledge embedded in the creator’s historical body of work.

\enlargethispage{\baselineskip}

\subsection{Comparative Evaluation Design}
Before the experiment, we formulated a stage-wise working hypothesis based on current generative AI capabilities. We assumed substantial reductions in pre-production and illustration time, but conservatively retained a large manual burden in Completion. Assuming that cross-page consistency repair and final watercolor-texture harmonization would remain largely human-dependent, we retained 83\% of the conventional Completion workload and computed the predicted Completion value as 603.3 $\times$ 0.83 $\approx$ 500.0 hours. This yielded a predicted total of approximately 637.2 hours, corresponding to an estimated 70.5\% reduction from the conventional 2,162.8-hour baseline.

We compare two complete picture-book productions by the same creator:
a conventional hand-drawn baseline (\textit{Carpe Diem}, 2020) and an AI-collaborative case (\textit{Golden Drops Opening the Sky}, 2026).
Production is decomposed into stages (planning, scenario, storyboard, sketch, illustration, finishing, layout, integration), and stage-wise time and substitution rates are analyzed; tool-development time is excluded from production-time accounting.

\begin{figure*}[!t]
  \centering
  \includegraphics[width=\textwidth]{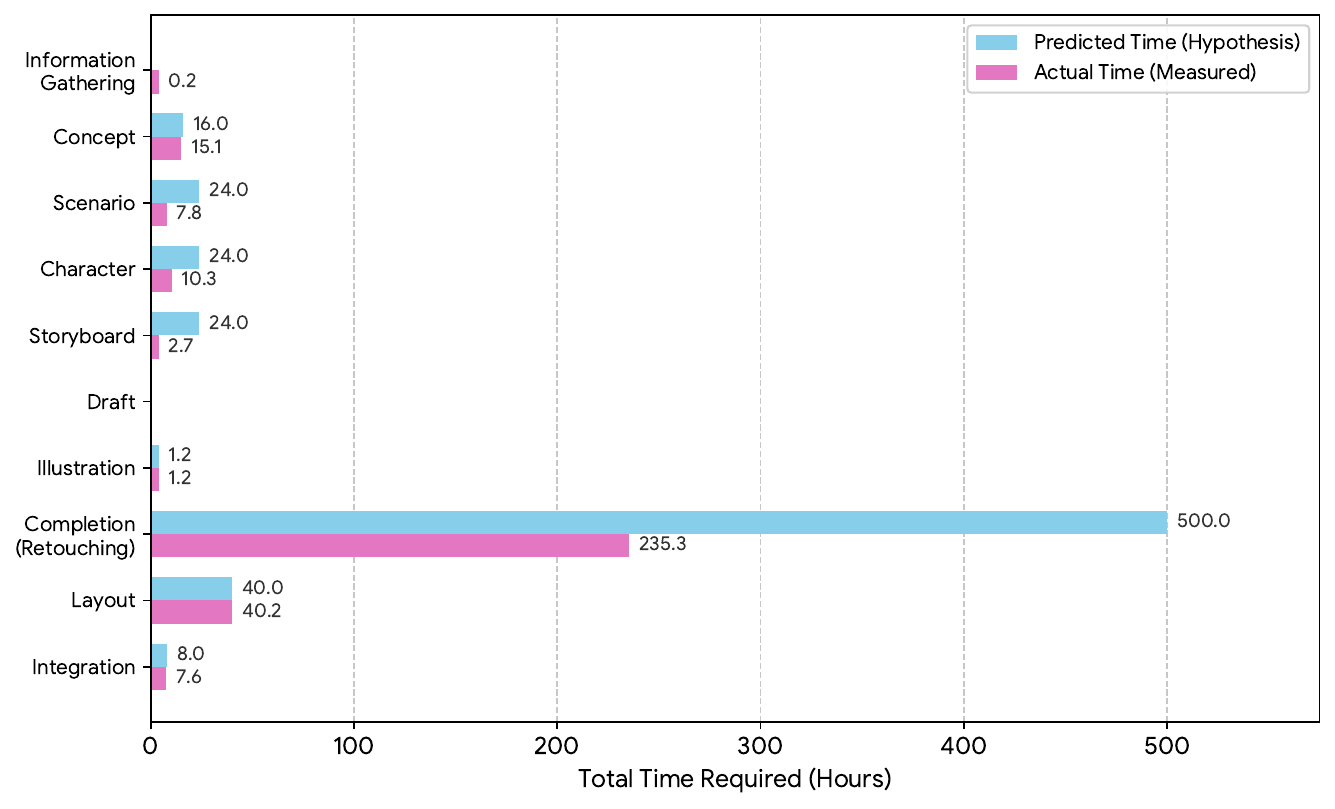}
  \caption{Predicted vs. actual time in the AI-collaborative pipeline. The chart shows stage-specific gaps between hypothesis and measured values, with Completion remaining the dominant time sink in the AI workflow.}
  \label{fig:predicted_observed_chart}
\end{figure*}

\section{Results and Discussion}
Quantitatively, the proposed workflow reduces total production time from 2,162.8 to 320.4 hours (85.2\%) and increases productivity by 575.0\%.
A per-title comparison between the conventional hand-drawn pipeline and the measured AI-collaborative pipeline is provided (see Figure~\ref{fig:before_actual}). The largest absolute reduction comes from the drafting-heavy illustration stage, which decreases from 1,203.6 hours to 1.2 hours (99.9\% substitution), while substantial human effort remains in post-generation integration.
Although the overall production time was reduced by 85\%, notable stage-level deviations from the hypothesis emerged. In the 'Completion' (retouching) phase, our preliminary estimate was 500.0 hours, whereas the observed value was 235.3 hours.
The stage-wise time breakdown is summarized (see Figure~\ref{fig:time_graph}).
In addition, predicted and observed time across finer-grained production stages are compared (see Figure~\ref{fig:predicted_observed_chart}); as the chart indicates, Completion remained the largest block of labor, even though its observed value was substantially lower than the conservative initial estimate.
The same pattern further indicates (see Figure~\ref{fig:predicted_observed_chart}) that prediction error is not uniform across stages: Concept, Layout, and Integration remained close to expectations, while Scenario, Character, and Storyboard required markedly less time than predicted.

While the use of LLMs significantly streamlined Scenario development (reducing labor from 41.2 to 7.8 hours), the Concept generation stage still required 15.1 hours of intensive human thought. This suggests that while AI can rapidly iterate narrative structures, the foundational philosophical core of a picture book remains a human-centric endeavor.

Yet the critical finding is how the recovered time is used.
Rather than terminating creative effort, the pipeline shifts labor toward high-value human processes:
aesthetic selection among many outputs and manual final integration for narrative coherence.
This shift is illustrated (see Figure~\ref{fig:completion_detail}), highlighting Completion-driven qualitative gains.

Importantly, the 235 hours of manual Completion are not mere error correction.
They constitute the decisive stage where local elements are integrated into a coherent scene through lighting balance, brush-level texture control, and expressive tuning.
The result is not homogenized automation; it is a production ecology in which human sensibility is amplified.

\begin{table}[t]
  \centering
  \caption{Breakdown of the 235 hours dedicated to Completion in the proposed workflow.}
  \label{tab:completion_breakdown}
  \begin{tabular}{@{}p{0.56\columnwidth}r@{}}
    \toprule
    Task Category & Hours \\
    \midrule
    Aesthetic Consistency (style unification across pages) & 72 \\
    Narrative Coherence (scene-level continuity repair) & 61 \\
    Facial/Emotional Refinement (expression micro-adjustments) & 44 \\
    Lighting and Atmospheric Calibration & 28 \\
    Background Density and Prop Integration & 18 \\
    Local Artifact Removal and Texture Harmonization & 12 \\
    \midrule
    \textbf{Total Completion Time} & \textbf{235} \\
    \bottomrule
  \end{tabular}
\end{table}

The largest share of Completion time is shown (see Table~\ref{tab:completion_breakdown}); it is devoted not to simple correction but to cross-page aesthetic and narrative integration. This distribution supports the claim that high-quality picture-book production still requires prolonged human synthesis after rapid AI drafting.

\subsection{The Qualitative Value of "Completion"}
\begin{figure*}[!t]
  \centering
  \includegraphics[width=1\textwidth]{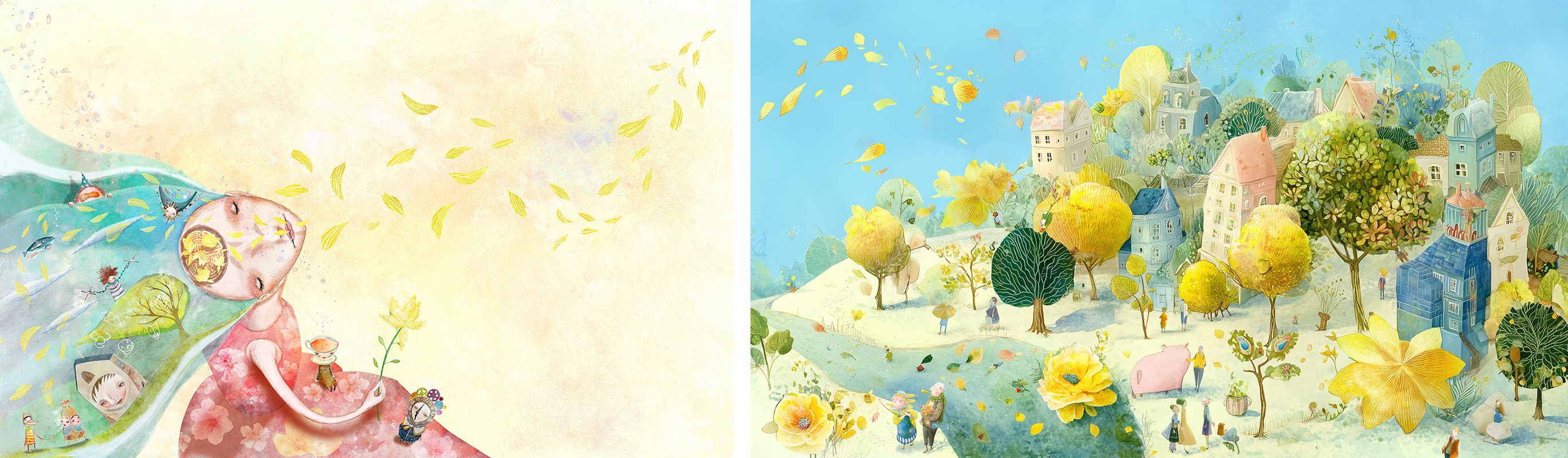}
  \caption{Expressive deepening through human Completion.
  (a) Conventional workflow.
  (b) Proposed AI-collaborative workflow.
  While initial drafts are AI-generated, 235 hours of embodied and aesthetic human intervention enable finer character depiction and substantially denser background articulation.
  The workflow allows creators to concentrate effort on qualitative values that are currently difficult to automate, including warm watercolor-like softness and emotionally calibrated color direction.}
  \label{fig:completion_detail}
\end{figure*}

While the quantitative reduction in production time is substantial (85\%), the core contribution of this workflow lies in the qualitative transformation enabled by what we term "Completion." As illustrated in the visual comparison (see Figure~\ref{fig:completion_detail}), raw AI-generated outputs, though structurally complex, often lack emotional resonance, spatial atmospheric presence, and stylistic consistency required for professional picture-books. The 235 hours dedicated to Completion are not merely for error correction; they represent an intensive process of aesthetic synthesis.

The gap between the estimated 500.0 hours and the observed 235.3 hours for Completion is informative. One possible interpretation is that the initial plan adopted a conservative assumption about retouching load, while the integrated AI-assisted drafting and retrieval workflow reduced corrective overhead more than expected. At the same time, the remaining 235.3 hours still indicate that the 'last mile' of artistic production---including narrative consistency and emotional resonance---depends on deliberate human intervention that current AI systems cannot fully replace.

With the rise of generative AI, image making is often misconstrued as becoming "magically easy." Our empirical results challenge that narrative. In the AI-collaborative workflow, the Completion stage (manual retouching and correction), which accounts for approximately 73\% of total time (235.3 hours), functioned as highly physical and technical \textit{Hard Work}: maintaining delicate watercolor textures while repairing generation-specific breakdowns and cross-scene inconsistencies.

Yet this physical difficulty did not map directly onto psychological depletion. Compared with the 2,162.8 hours of fully manual production, experienced as solitary agony, the interpretive and restorative engagement with unpredictable AI-generated drafts was perceived instead as dialogic co-creation with AI---a form of \textit{Mild Work} (labor accompanied by intrinsic enjoyment and a sense of accomplishment).

This paradox suggests that AI does not necessarily deprive creators of the joy of creation. Rather, by offloading repetitive rendering labor, it can reallocate user effort toward higher-order aesthetic judgment and embodied retouching, thereby redefining and potentially improving the creative user experience (UX).

Through embodied manual intervention, the creator meticulously curates lighting and shadows, refines the subtle nuances of character expressions, and orchestrates the textural coherence of the brushwork. This manual phase explicitly addresses the inherent visual hallucinations and structural artifacts typical of diffusion models, removing disjointed elements to ensure strict narrative continuity across the 15 illustrations. This physical engagement echoes foundational design theories which posit that creative professionals "think in action" through direct dialogue with their materials \cite{schon1983, pallasmaa2009}, reasserting the value of the "practiced digital hand" \cite{mccullough1998}.

Additionally, Completion injects "lyricism" and "density" into the scene. The creator adjusts the contextual weight of props and architecture, transforming them from mere decorative background elements into active narrative affordances, a critical requirement in picture-book theory where visual details must silently carry the weight of the story \cite{nodelman1988, nikolajeva2001, shulevitz1997}. By doing so, what begins as an inorganic assembly of generated pixels is woven into a meaningful, emotionally cohesive narrative space. Thus, the traditional act of drawing appears to be not eliminated but reconfigured: the creator's agency shifts from the mechanical execution of lines to the high-level orchestration of aesthetic and emotional value.

\subsection{Why Completion Required 235 Hours: A Three-Axis Interpretation}
The Completion workload can be interpreted as \textit{last-mile creativity}, in which quality-critical decisions concentrate near the end of the pipeline. We organize this gap between predicted and observed effort into three analytical axes:

\paragraph{Aesthetic Consistency}
Even when initial generation is stylistically constrained, page-to-page drift persists in brush texture, edge softness, and color temperature. Maintaining a coherent authorial signature across 15 illustrations required repetitive but Judgment-intensive interventions to align micro-level rendering choices with the macro-level visual identity of the book.

\paragraph{Narrative Coherence}
Diffusion-based outputs can produce locally plausible details that are globally inconsistent with story logic. Substantial time was therefore spent reconciling object continuity, directional lighting logic, and scene transitions so that visual cues support the narrative arc rather than fragment it.

\paragraph{Cognitive Load}
The pipeline reduced physical drawing repetition, but shifted burden toward cognitively intensive comparison and decision-making. Instead of prolonged motor execution, the creator performed sustained evaluative work across many candidate outputs, balancing stylistic fit, emotional tone, and narrative function. This shift from physical fatigue to cognitive load helps explain why Completion remains time-intensive in expert workflows.

\vspace{0.8\baselineskip}
\subsection{Limitations and Future Work}
This study is based on one in-depth professional case (one picture-book title with 15 illustrations), and the findings should therefore be interpreted as analytically rich but statistically bounded evidence. In particular, the measured substitution rates and Completion workload may vary across creators with different stylistic priors, production habits, and tolerance for visual imperfection.

A key next step is cross-creator validation of the pipeline components introduced here, especially the \textit{Prompt Recipe} mechanism for style inheritance and the metadata automation chain (Gemini-based tagging + thesaurus normalization + XMP/IPTC writing). Evaluating these tools with multiple illustrators would clarify which parts are universally transferable and which require adaptation to individual tacit practices.

Future work should also extend the protocol to longer narrative formats, such as multi-volume picture-book projects or hybrid editorial pipelines combining still imagery and animation. Such settings would allow direct testing of whether asynchronous generation management and retrieval-ready metadata produce compounding gains at larger scales.

Finally, even as generative models continue to improve, we expect the value of embodied human intervention in Completion to remain structurally important. Higher-fidelity generation may reduce technical correction, but aesthetic commitment, narrative responsibility, and emotionally calibrated final decisions are likely to remain human-centered functions in professional storytelling workflows.

\vspace{-0.6\baselineskip}
\enlargethispage{2\baselineskip}
\section{Conclusion}
This study provides an empirical evaluation of an AI-collaborative workflow for professional picture-book production, showing an 85.2\% reduction in total time (2,162.8 to 320.4 hours). The key contribution, however, is not speed alone: delegating early drafting to AI reallocates human effort to high-level \textit{Judgment} and embodied \textit{Completion}.

Although AI outputs are structurally strong, publication-quality results still require substantial aesthetic synthesis. The observed 235 hours of Completion indicate that the act of drawing is reconfigured rather than removed, preserving the creator's role in emotional calibration, narrative coherence, and stylistic identity \cite{schon1983, epstein2023}.

Accordingly, we frame generative AI as a high-throughput instrument within a human-centered production ecology: technology accelerates exploration, while authorship remains grounded in human intervention. Generalizability limits of this single-case study are detailed in the Limitations and Future Work subsection.

\vspace{0.8\baselineskip}
\section*{Acknowledgments}
This research was supported by VoiceCast Co., Ltd. The author would like to thank Kai Hosoya for his invaluable collaboration during the production of "Golden Drops Opening the Sky."
The author also expresses sincere gratitude to Haruna Fujita and Moeka Muramatsu (Shizuoka University of Art and Culture) for their substantial support in scanning picture-book materials and their important assistance in the AI-based illustration-to-video process, respectively.

\newpage
\bibliographystyle{unsrturl}
\bibliography{references}

\end{document}